\documentclass[12pt]{article}
\usepackage{graphicx}
\usepackage[numbers,square]{natbib}

\usepackage[left=2cm,right=2cm,top=2.5cm,bottom=2.5cm]{geometry}

\usepackage{graphicx}
\usepackage{subcaption}
\usepackage{physics}
\usepackage{bigints}
\usepackage{braket}
\usepackage{tikz-feynman}
\usepackage{float}
\usepackage{mathtools}
\usepackage{xcolor}
\usepackage{cancel}
\usepackage{caption}
\usepackage{feynmf} 
\usepackage{amssymb}
\usepackage{amsfonts}
\usepackage{epsf}
\usepackage{epsfig}
\usepackage{rotating}
\usepackage{graphicx}
\usepackage{amsmath}
\usepackage{fancyhdr}
\usepackage{graphics}
\usepackage{color}
\usepackage{frontespizio}

\usepackage{type1ec}
\usepackage[T1]{fontenc}
\usepackage{lettrine}
\usepackage{bbold}
\usepackage{calligra}
\usepackage{tikz}
\usepackage{mathrsfs}
\usepackage{curve2e}
\usepackage{setspace}
\usepackage{indentfirst}
\usepackage{emptypage}
\usepackage[babel]{csquotes} 
\usepackage[font=small,labelfont=bf,labelsep=quad]{caption} 
\usepackage{graphicx} 
\usepackage{listings} 
\usetikzlibrary{patterns}
\usepackage{relsize}
\usetikzlibrary{intersections,positioning}
\usetikzlibrary{decorations.pathmorphing,decorations.markings,arrows,positioning}
\usepackage{braket}
\usepackage{mathrsfs}
\usepackage{stackengine}
\usepackage{calc}
\newlength\shlength
\newcommand\xshlongvec[2][0]{\setlength\shlength{#1pt}%
	\stackengine{-5.6pt}{$#2$}{\smash{$\kern\shlength%
			\stackengine{7.55pt}{$\mathchar"017E$}%
			{\rule{\widthof{$#2$}}{.57pt}\kern.4pt}{O}{r}{F}{F}{L}\kern-\shlength$}}%
	{O}{c}{F}{T}{S}}

\usepackage{hyperref}
\hypersetup{
	colorlinks,
	citecolor=black,
	filecolor=black,
	linkcolor=black,
	urlcolor=black
}

\IfFileExists{dsfont.sty}
{\usepackage{dsfont}
	\let\mathbb=\mathds
	\newcommand{\id}{\mathds{1}}}
{\typeout{Package dsfont.sty was not found, using alternative macros.}
	\let\mathds=\mathbb
	\newcommand{\id}{\mbox{1 \kern-.59em {\rm l}}}}

\usepackage{slashed}
\usepackage{units}
\usepackage{setspace}
\let\a=\alpha   \let\b=\beta   \let\g=\gamma   \let\d=\delta
 \let\z=\zeta        
      \let\l=\lambda  \let\m=\mu
\let\n=\nu                 \let\r=\rho
\let\s=\sigma  \let\t=\tau

\let\d=\delta
\let\s=\sigma
\renewcommand{\a}{\alpha}

\renewcommand{\r}{\rho}
\renewcommand{\t}{\tau}

\def\nbox#1#2{\vcenter{\hrule \hbox{\vrule height#2in
			\kern#1in \vrule} \hrule}}
\def\sq{\,\raise.5pt\hbox{$\nbox{.09}{.09}$}\,}
\def\sqb{\,\raise.5pt\hbox{$\overline{\nbox{.09}{.09}}$}\,}

\newcommand{\bea}{\begin{eqnarray}}
\newcommand{\eea}{\end{eqnarray}}
\newcommand{\be}{\begin{equation}}
\newcommand{\ee}{\end{equation}}

\newcommand{\bes}{\begin{subequations}}
	\newcommand{\ees}{\end{subequations}}

\newcommand{\brak}[1]{\left({#1}\right)}
\numberwithin{equation}{section}

\usepackage{accents}

\makeatletter
\newcommand{\xLine}[2][]{\ext@arrow 0359\Rightarrowfill@{#1}{#2}}
\makeatother
\xdefinecolor{darkgreen}{RGB}{102, 204, 70}
\xdefinecolor{darkblue}{RGB}{0, 0, 153}

\usepackage{xcolor}

\usepackage{comment}
\title{331 Review}
\author{Dario MELLE}
\date{February 2024}

\begin{document}
\vspace{1.5cm}
\begin{center}

{\bf\Large Heavy Quark Decays in the Bilepton Model}
  \vspace{0.2cm}
{\bf  \Large }


 \vspace{0.3cm}
\vspace{1cm}
{\bf $^{(1)}$Gennaro Corcella, $^{(2,3)}$Claudio Corian\`o, $^{(2,3)}$Paul H. Frampton and $^{(2,3)}$Dario Melle\\}

\vspace{1cm}

{\it
	$^{(1)}$INFN, Laboratori Nazionali di Frascati, Via E. Fermi 54, 00044, Frascati (RM), Italy
	 $^{(2)}$Dipartimento di Matematica e Fisica, Universit\`{a} del Salento \\
	and INFN, Sezione di Lecce, Via Arnesano, 73100, Lecce, Italy\\
	$^{(3)}$Centro Nazionale di Ricerca in High Performance Computing,
        Big Data\\ and Quantum Computing,
Via Magnanelli 2, 40033, Casalecchio di Reno (BO), Italy}
\vspace{0.5cm}
\end{center}

\begin{abstract}
 Given the current absence of new physics signals at the LHC, it is increasingly important to investigate alternative scenarios beyond those commonly explored. In this work, we study a variant of the 331 model that predicts the existence of vector bileptons with electric charge and lepton number $\pm 2$, as well as TeV-scale exotic quarks carrying charges $\pm 5/3$ and $\pm 4/3$. Specifically, we focus on the primary production of heavy quarks with charge $\pm 5/3$, which decay into a bottom quark and a bilepton, followed by the bilepton's decay into same-sign muon pairs. As a case study, we select a benchmark point that complies with current experimental exclusion limits and theoretical expectations for the bilepton mass. Our analysis shows that the resulting signal stands out clearly from Standard Model backgrounds and could be observed at a future 100 TeV hadron collider such as FCC-$hh$. In contrast, the LHC, even in its high-luminosity phase, lacks the sensitivity required to detect this signal.
\end{abstract}

\newpage
\section{Introduction}
The search for new physics at the Large Hadron Collider (LHC) has revealed persistent gaps in our understanding of fundamental forces, particularly regarding the limitations of the Standard Model (SM). Two of the most pressing issues are the hierarchy problem in the Higgs sector and smallness of the
neutrino masses. To address these challenges, theorists often propose extensions of the SM that include larger gauge groups and additional particles. Grand Unified Theories (GUTs) are a prominent example, offering a unified framework for the forces. However, they operate at energy scales, approximately from
$10^{14}$ to $10^{16}$ GeV, which are far beyond the reach of the LHC and the future colliders
currently under debate, such as  the hadron machine FCC-$hh$ foreseen at 100 TeV.
Bridging this vast energy gap down to the TeV scale, which is accessible to the LHC and ultimately
FCC-$hh$, remains a major challenge due to the complexity and richness of these extended theories.

Nevertheless, certain models predict that signs of new physics could still emerge at the LHC
or FCC-$hh$ scales, particularly in scenarios involving enlarged gauge symmetries. One such model is the bilepton model \cite{framptonChiralDileptonModel1992,pisanoSU3U1Model1992}, based on the gauge group $SU(3)_C \times   SU(3)_L \times   U(1)_X$, also known as 331 model. This scenario is particularly attractive because it provides solutions to several shortcomings of the SM, including explanations for neutrino masses and charge quantization, while remaining testable at TeV energies. 
This reshuffling necessitates that, already at TeV scale, the three families transform differently under the gauge group; this can explain the failure of $SU(5)$ Grand Unification Theory, where it was assumed that all three families remain sequential up to the GUT scale $\sim 10^{11}-10^{13}$ TeV.
The bilepton model introduces the new gauge group $SU(3)_L$, an extension analogous to the familiar colour gauge group $SU(3)_C$ of QCD, and reorganizes quarks and leptons into multiplets of $SU(3)_L$,
with fermions transforming as triplets or antitriplets, depending on their charge and quantum
properties. This reshuffling of particle content leads to new interactions and potentially observable
phenomena even at current collider energies.

A hallmark of the 331 model is the prediction of novel gauge bosons, known
as bileptons and labelled as $Y^{++/--}$ hereafter, which carry both lepton number and electric charge. These particles, with charges $Q = \pm2$ and
lepton number $L=\pm2$ , open up a new avenue for exploring physics beyond the SM. Bileptons can lead to distinctive signatures, such as lepton-flavour violating processes or new decay channels for heavy particles, which could be detected at the LHC or future colliders.
Spontaneous symmetry breaking features additional complexities: it occurs at an energy scale close to the TeV region, transitioning the gauge group down to the familiar electroweak symmetry. This process
generates the masses of the gauge bosons, including the SM-like $W^\pm$ and $Z^0$, as well as the
new bileptons.

The 331 model provides a framework for addressing key open questions, such as the origin of neutrino masses and the unification of forces at high energies. It also constrains its parameter space through the requirement of real gauge couplings, which limits possible signals and enhances its predictive power. Despite these promising features, the model still faces experimental scrutiny and requires further theoretical refinements to assess its full viability. In the specific realization of the 331 model that includes bileptons, symmetry breaking occurs through Higgs fields whose vacuum expectation values (vevs) are tied to the TeV scale. This process not only gives mass to the exotic bileptons, but also introduces a rich phenomenology of new particles, such as the doubly-charged gauge bosons $Y^{++}$ and $Y^{--}$.
Furthermore, at energies beyond 14 TeV, but accessible to the 100 TeV  FCC-$hh$, QCD will possibly bind 
the new quarks predicted by the bilepton model and produce a plethora of additional baryons and mesons \cite{framptonAdditionalBaryonsMesons2021}.
Overall, the bilepton version of the 331 model stands out as a compelling extension of the SM, with potential observable consequences in high-energy experiments. It offers a unique window into new physics, linking low-energy phenomena to high-energy symmetries. Whether the LHC, including the
high-luminosity HL-LHC, can uncover evidence for bileptons or other exotic particles will play a crucial role in determining the future direction of particle physics beyond the Standard Model.

This work completes a series of papers on the theory and phenomenology of the bilepton model.
In detail,
in Refs.~\cite{corcellaBileptonSignaturesLHC2017,corcellaExploringScalarVector2018}
we investigated bilepton-pair production and decay into same-sign lepton pairs, such as
$Y^{++}\to \mu^+\mu^+$, with
Ref.~\cite{corcellaExploringScalarVector2018} also exploring possible
differences between signals
of vector and scalar bileptons. The conclusion of
Refs.~\cite{corcellaBileptonSignaturesLHC2017,corcellaExploringScalarVector2018} was that, for a bilepton mass about 900 GeV, consistent with the
exclusion limits at that time, it would have been possible to discover bileptons at LHC,
with the vector signal greatly overwhelming the scalar one.
More recently, Ref.~\cite{corcellaNonleptonicDecaysBileptons2022} explored possible nonleptonic decays
of bileptons, namely decays into a heavy TeV-scale quark plus a Standard Model heavy quark, and found
that such a signal will potentially be visible at the future FCC-$hh$, as the LHC statistics are
too low, even in the high luminosity phase.
Ref.~\cite{corcellaNonleptonicDecaysBileptons2022} assumed a bilepton mass
around 1.3 TeV, consistently with the findings, based on renormalization group
arguments, of Ref.~\cite{Coriano:2020iiz}.
We underline that, regardless of the decay modes, a general
result of Refs.~\cite{corcellaBileptonSignaturesLHC2017,corcellaExploringScalarVector2018,corcellaNonleptonicDecaysBileptons2022} is that, given the high mass of the bilepton and consequently
of the lepton or heavy-quark pairs produced in the decay, it is relatively straightforward
to discriminate any bilepton signal from the Standard Model backgrounds.
To our knowledge, there is currently no specific experimental analysis on
vector bileptons, while both ATLAS and CMS have searched for doubly-charged
scalar Higgs bosons in some scenarios beyond the Standard Model.
In detail, ATLAS \cite{ATLAS:2022pbd} has set the limit
$m_{H^{\pm\pm}}>1080$~GeV in left-right symmetric type-II see-saw model
$m_{H^{\pm\pm}}>900$~GeV in the Zee--Babu neutrino-mass model at
$\sqrt{s}=13$~TeV and integrated luminosity ${\cal L}=139$~fb$^{-1}$.
CMS set instead limits between 535 and 820 GeV according to the
lepton flavours for leptonic doubly-charged Higgs decays
at 13 TeV and 12.9 fb$^{-1}$ \cite{CMS:2017pet}.

In the present paper we study
primary production of heavy quarks, such as $T^\pm$, charged $\pm 5/3$, and their subsequent decays into
bileptons and Standard Model quarks, with the bileptons again decaying into same-sign leptons. To our knowledge, an analysis on such heavy quarks focused
on the 331 model has never been carried out so far.
Nevertheless, a companion investigation was undertaken in \cite{corcellaVectorlikeQuarksDecaying2021}, where the authors, in the framework of
simplified models,  studied pair production of vector-like heavy quarks
with charge 5/3 and decay into singly- or doubly-charged bosons plus a
top or a bottom quark. A CMS analysis searching for top partners decaying to a top quark and a $W$
boson \cite{CMS:2020ttz} was also recast in Ref.~\cite{corcellaVectorlikeQuarksDecaying2021}
and the results were eventually provided as recast
efficiencies, which allow reinterpretation in any model.
More recently, Ref.~\cite{Liu:2024hvp} explored single production
of heavy quarks with charge 5/3 at the LHC and presented exclusion
limits and discovery reach, typically
in the range of few TeV, for a few values of the integrated luminosity.
From the experimental viewpoint, such heavy quarks, labelled as
$X_{5/3}$, were searched for
in Ref.~\cite{CMS:2018ubm} by the CMS collaboration at $\sqrt{s}=13$~TeV
and integrated luminosity ${\cal L}=35.9$~fb$^{-1}$ and lower limits at 95\%
confidence level on their mass were set at  1.33 and 1.30~TeV
for right-handed and left-handed couplings to $W$ bosons, respectively.
Furthermore, vector-like quarks as well as
singly and  doubly-charged gauge bosons
in the minimal 331 model
were also explored in \cite{Calabrese:2023ryr}, where the authors
considered both single and pair production of bileptons, possibly
accompanied by one extra jet. The results were then compared with the
ATLAS searches for a doubly-charged Higgs boson \cite{ATLAS:2022pbd}.

The present paper focuses on heavy quarks in the bilepton model and is organized
as follows.
In Section 2 we review the particle content of the minimal 331 model, while in Section 3 we scrutinize
Higgs and Yukawa sectors. Section 4 will discuss the choice of the benchmark point for the analysis
and Section 5 present some phenomenological results, mostly at FCC-$hh$.
Section 6 will contain some concluding remarks.

\section{The particle content of the minimal 331 Model}\label{331}

As discussed in the introduction, the 331 model is 
based on a $SU(3)_C \times   SU(3)_L \times   U(1)_X$ gauge symmetry.
In its minimal version,
the symmetry group $SU(3)_L$ requires the addition of three exotic quarks to the Standard Model quark content. 
The first two generations of quarks are placed in triplets of $SU(3)_L$ along with exotic quarks $D$ or $S$: 
\begin{equation}
	\begin{pmatrix}
		u\\d\\ D
	\end{pmatrix}_L,
	\begin{pmatrix}
		c\\s\\ S
	\end{pmatrix}_L\rightarrow \brak{\textbf{3},\textbf{3},-\frac{1}{3}}\ ,
\end{equation}
with a quantum number $X=-\frac{1}{3}$ under $U(1)_X$.
In contrast, the third generation is placed in an anti-triplet: \begin{equation}
	\begin{pmatrix}
		b\\-t\\ T
	\end{pmatrix}_L
	\rightarrow \brak{\textbf{3},\bar{\textbf{3}},\frac{2}{3}}\ ,
\end{equation}
with $X=\frac{2}{3}$.
Conversely, leptons in each of the three families are treated uniformly as $SU(3)_L$ anti-triplets, colour singlets, represented as: 
\begin{equation}
	\begin{pmatrix}
		e\\-\n_e\\ e^c
	\end{pmatrix}_L,
	\begin{pmatrix}
		\m\\-\n_\m\\ \m^c
	\end{pmatrix}_L,
	\begin{pmatrix}
		\t\\-\n_\t\\ \t^c
	\end{pmatrix}_L\rightarrow \brak{\textbf{1},\bar{\textbf{3}},0}\ ,
\end{equation}
 with $X=0$. Here, $e^c$, $\mu^c$, and $\tau^c$ are the left-handed Weyl spinors of the charge conjugate fields.

For each left-handed quark field, there is a corresponding right-handed singlet under $SU(3)_L$ with the following quantum numbers:\begin{equation}
	\begin{matrix}
		(u^c)_L\\
		(c^c)_L\\ 
		(t^c)_L
	\end{matrix} \rightarrow \brak{\bar{\textbf{3}},\textbf{1},-\frac{2}{3}}\ ,
	\ \ \ \ \ \ \ \ \ 	\begin{matrix}
		(d^c)_L\\
		(s^c)_L\\
		(b^c)_L
	\end{matrix} \rightarrow \brak{\bar{\textbf{3}},\textbf{1},\frac{1}{3}}\ ,
\end{equation}

\begin{equation}
	\begin{matrix}
		(D^c)_L\\(S^c)_L
	\end{matrix} \rightarrow \brak{\bar{\textbf{3}},\textbf{1},\frac{4}{3}}\ ,
	\ \ \ \ \ \ \ \ \ 
	\begin{matrix}
		(T^c)_L
	\end{matrix} \rightarrow \brak{\bar{\textbf{3}},\textbf{1},-\frac{5}{3}}\ ,
\end{equation}

The $U(1)_X$ charges for the first two families are $-\frac{2}{3}$, $\frac{1}{3}$, and $\frac{4}{3}$ for $u^c$, $d^c$, and $D^c$, respectively, with identical charges for the second generation. For the third generation, the $X$ quantum numbers are $\frac{1}{3}$, $-\frac{2}{3}$, and $-\frac{5}{3}$ for $b^c$, $t^c$, and $T^c$, respectively.

Three scalar triplets under $SU(3)_L$ are introduced to trigger spontaneous symmetry breaking:
\begin{equation}
	\r=\begin{pmatrix}
		\r^{++}\\ \r^+\\ \r^0
	\end{pmatrix}\ ,\ 
	\eta=\begin{pmatrix}
		\eta^{+}_1\\ \eta^0\\ \eta^-_2
	\end{pmatrix}\ ,\ 
	\chi=\begin{pmatrix}
		\chi^{0}\\ \chi^-\\ \chi^{--} 
	\end{pmatrix} \ ,
\end{equation}
with charges $X=1,0,-1$, respectively. Additionally, a scalar sextet is necessary for generating physical lepton masses: 
\begin{equation}
	\s=\begin{pmatrix}
		\s^{++}_1&\frac{\s_1^+}{\sqrt{2}}&\frac{\s_1^0}{\sqrt{2}}\\
		\frac{\s_1^+}{\sqrt{2}}&\s_2^0&\frac{\s^-_2}{\sqrt{2}}\\ 
		\frac{\s_1^0}{\sqrt{2}}&\frac{\s^-_2}{\sqrt{2}}&\s_2^{--}\\
	\end{pmatrix}\ ,
\end{equation}
This unconventional assignment of quantum numbers ensures that gauge anomalies are not cancelled individually for each family, as in the Standard Model. Instead, the contributions from all quarks must be combined to achieve anomaly cancellation, which is one of the key features of the model. This mechanism offers a potential explanation for the existence of three generations of fermions, providing insight into the so-called flavour puzzle and guiding the development of new models in particle physics. The interfamily anomaly cancellation requires the number of families $n_f$  to be a multiple of the number of the QCD colours $(N)$, $n_f= k \, N$, with $k$ an integer, 
but asymptotic freedom of QCD requires $k=1$ and hence $n_f=3$
 exactly.

\section{Higgs Sector}
The particle content of the spectrum of the 331 model is very rich due to the presence of three scalar triplets and a scalar sextet. In fact, spontaneous
symmetry breaking happens in two steps. In the first one, namely $SU(3)_L\times   U(1)_X \to SU(2)_L\times   U(1)_Y$, only the scalar triplet $\r$ acquires
a vacuum expectation value (vev), i.e.  
\begin{equation}
	\expval{\r}=\begin{pmatrix}
		0\\0\\\frac{v_\r}{\sqrt{2}}
	\end{pmatrix}\ ,
\end{equation}
while the other scalars get a vev only in the subsequent breaking $SU(2)_L\times   U(1)_Y\to   U(1)_{\rm{em}}$, i.e.
\begin{equation}
	\expval{\eta}=\begin{pmatrix}
		0\\ \frac{v_\eta}{\sqrt{2}}\\0
	\end{pmatrix} \ ,\ \ \ \
	\expval{\chi}=\begin{pmatrix}
		\frac{v_\chi}{\sqrt{2}}\\0\\0
	\end{pmatrix} \ ,\ \ \ \
	\expval{\s}=\begin{pmatrix}
		0&0&\frac{v_\s}{2}\\
		0&0&0\\
		\frac{v_\s}{2}&0&0
	\end{pmatrix}\ .
\end{equation}
In principle, also the component $\s_2^0$ can acquire a vev, a choice which
is particularly relevant for neutrino masses \cite{tullyGeneratingNeutrinoMass2001}, and hence contributes to enlarge the phenomenology of the model; in
the following, we shall instead set the vev of $\s_2^0$ to zero.
After electroweak symmetry breaking (EWSB), in the scalar sector one has five scalar neutral  Higgses, three pseudoscalar Higgses, four charged Higgses and three doubly-charged Higgses. The (lepton-number conserving) potential of the model is given by  \cite{tullyScalarSector3312003}:
\allowdisplaybreaks
\begin{align}\label{pot}
	V=&m_1\r^\dagger\r+m_2\eta^\dagger\eta+m_3\chi^\dagger\chi+\l_1(\rho^\dagger\r)^2+\l_2(\eta^\dagger\eta)^2+\l_3(\chi^\dagger\chi)^2\nonumber\\
	&+\l_{12}\r^\dagger\r\eta^\dagger\eta+\l_{13}\r^\dagger\r\chi^\dagger\chi+\l_{23}\chi^\dagger\chi\eta^\dagger\eta+\z_{12}\r^\dagger\eta\eta^\dagger\r+\z_{13}\r^\dagger\chi\chi^\dagger\r+\z_{23}\eta^\dagger\chi\chi^\dagger\eta\nonumber\\
	&+m_4\Tr(\s^\dagger\s)+\l_4(\Tr(\s^\dagger\s))^2+\l_{14}\r^\dagger\r\Tr(\s^\dagger\s)+\l_{24}\eta^\dagger\eta\Tr(\s^\dagger\s)+\l_{34}\chi^\dagger\chi\Tr(\s^\dagger\s)\nonumber\\
	&+\l_{44}\Tr(\s^\dagger\s\s^\dagger\s)+\z_{14}\r^\dagger\s\s^\dagger\rho+\z_{24}\eta^\dagger\s\s^\dagger\eta+\z_{34}\chi^\dagger\s\s^\dagger\chi\nonumber\\
	&+(\sqrt{2}f_{\r\eta\chi}\epsilon^{ijk}\r_i\eta_j\chi_k+\sqrt{2}f_{\r\s\chi}\r^T\s^\dagger\chi\nonumber\\
	&+\xi_{14}\epsilon^{ijk}\r^{*l}\s_{li}\r_j\eta_k+\xi_{24}\epsilon^{ijk}\epsilon^{lmn}\eta_i\eta_l\s_{jm}\s_{kn}+\xi_{34}\epsilon^{ijk}\chi^{*l}\s_{li}\chi_j\eta_k )+\textnormal{h.c.}\ .
\end{align}
This potential can be extended by incorporating additional lepton-number violating terms that are singlets under the 331 gauge group, as discussed in detail in \cite{tullyScalarSector3312003} and further elaborated in \cite{fonsecaLeptonNumberViolation2016}. The EWSB mechanism induces a mixing between the Higgs fields, as shown in \cite{costantiniTheoreticalConstraintsHiggs2020}. From Eq. (\ref{pot}), one can derive explicit expressions for the mass matrices of scalar, pseudoscalar, charged, and doubly-charged Higgs bosons using standard techniques.
In the broken Higgs phase, the minimization conditions \be\label{mincond} \frac{\partial V}{\partial v_\phi}=0, \quad \langle \phi^0\rangle=v_\phi, \quad \phi=\rho, \eta, \chi, \sigma \ee define the tree-level vacuum structure. One-loop corrections to vacuum stability have been recently analyzed in
\cite{Dorsch:2024ddk}, for a simpler version of the model, known as the `economical' 331 model \cite{footSU3_LU1_NSU4_L1994,ponceANALYSISSU3cSU3LU1X2002}.

As pointed out above, in this work we consider massless neutrinos by setting the vacuum expectation value (vev) of the neutral field $\sigma_2^0$ to zero. This choice can be generalized to generate small Majorana neutrino masses, as suggested in \cite{tullyGeneratingNeutrinoMass2001}.
The explicit expressions of the minimization conditions are then given by:
\begin{align}\label{minpot1}
	m_1 v_\rho + \lambda_1 v_\rho^3 + \frac{1}{2}\lambda_{12}v_\rho v_\eta^2-f_{\rho\eta\chi} v_\eta v_\chi+\frac{1}{2}\lambda_{13}v_\rho v_\chi^2 - \frac{1}{\sqrt2}\xi_{14}v_\rho v_\eta v_\sigma + f_{\rho\sigma\chi}v_\chi v_\sigma&\\
	+ \frac{1}{2}\lambda_{14}v_\rho v_\sigma^2 + \frac{1}{4}\zeta_{14}v_\rho v_\sigma^2&=0\ ,\nonumber\\
	m_2 v_\eta + \frac{1}{2}\lambda_{12}v_\rho^2 v_\eta +\lambda_2 v_\eta^3 - f_{\rho\eta\chi} v_\rho v_\chi +\frac{1}{2}\lambda_{23} v_\eta v_\chi^2 - \frac{1}{2\sqrt2}\xi_{14}v_\rho^2 v_\sigma+ \frac{1}{2\sqrt2}v_\chi^2 v_\sigma&\\
	+\frac{1}{2}\lambda_{24} v_\eta v_\sigma^2-\xi_{24} v_\eta v_\sigma^2&=0\ ,\nonumber\\
	m_3 v_\chi + \lambda_3 v_\chi^3 + \frac{1}{2} \lambda_{13} v_\rho^2 v_\chi - f_{\rho\eta\chi} v_\rho v_\eta +\frac{1}{2}\lambda_{23}v_\eta^2 v_\chi +\frac{1}{\sqrt2}\xi_{34}v_\eta v_\chi v_\sigma + f_{\rho\sigma\chi} v_\rho v_\sigma&\\
	+\frac{1}{2}\lambda_{34} v_\chi v_\sigma^2 + \frac{1}{4}\zeta_{34} v_\chi v_\sigma^2&=0\ ,\nonumber\\
	\label{minpot2}
	m_4 v_\sigma + \frac{1}{2}\lambda_{14}v_\rho^2 v_\sigma + \lambda_{44} v_\sigma^3 + \frac{1}{2}\lambda_4 v_\sigma^3 + f_{\rho\sigma\chi} v_\rho v_\chi - \frac{1}{2\sqrt2} \xi_{14} v_\rho^2 v_\eta + \frac{1}{2\sqrt2} \xi_{34} v_\eta v_\chi^2&\\
	 + \frac{1}{4}\zeta_{14} v_\rho^2 v_\sigma + \frac{1}{2}\lambda_{24} v_\eta^2 v_\sigma - \xi_{24} v_\eta^2 v_\sigma + \frac{1}{2} \lambda_{34} v_\chi^2 v_\sigma + \frac{1}{4} \zeta_{34} v_\chi^2 v_\sigma&=0.
	\nonumber
\end{align}
These conditions are inserted into the the tree-level mass matrices of the CP-even and CP-odd Higgs sectors, derived from $M_{ij}=\left.{\partial^2 V}/{(\partial \phi_i\partial \phi_j)}\right|_{vev}$, where $V$ is the potential in
Eq.~(\ref{pot}). The mass eigenstates are defined as follows:
\begin{equation}
	h=R^S \begin{pmatrix}
		{\rm Re}\, \rho^0\\ {\rm Re}\, \eta^0\\ {\rm Re}\, \chi^0 \\ {\rm Re}\, \sigma^0_1\\ \s^{0}_2
	\end{pmatrix}\,,\ \ \ 
	Ah=R^P   \begin{pmatrix}
		{\rm Im}\, \rho^0\\ {\rm Im}\, \eta^0\\ {\rm Im}\, \chi^0 \\ {\rm Im}\, \sigma^0_1
	\end{pmatrix}\,,\ \ \ 
	H^+=R^C \begin{pmatrix}
		\r^{+}\\ \chi^{+}\\ \eta^{+}_1\\ \eta^{+}_2\\ \s^{+}_1 \\ \s^{+}_2
	\end{pmatrix}\,,\ \ \ 
	H^{++}=R^{2C} \begin{pmatrix}
		\r^{++}\\ \chi^{++}\\ \s^{++}_1 \\ \s^{++}_2
	\end{pmatrix}\,,
\end{equation}
where the explicit expressions of the mass matrices
are too cumbersome to be presented here, and are given in \cite{corianoSU3CSU3LU1X2024}.

As anticipated before, in this scenario we have five scalar Higgs bosons, one of which should correspond to the Standard Model Higgs with a mass of approximately 125 GeV. Additionally, there are four neutral pseudoscalar Higgs bosons:
two of them are the Nambu-Goldstone (NG) bosons associated with
the massive vector bosons $Z$ and $Z'$, while the other two will be labelled as $Ah_1$ and $Ah_2$.
Moreover, the spectrum includes six charged Higgs bosons ($H_1^\pm$, $\dots$,
$H_6^\pm$), two of which are charged NG bosons, along with three doubly-charged Higgses ($H_1^{\pm\pm}$, $H_2^{\pm\pm}$, $H_3^{\pm\pm}$),
one of which is also a NG boson.

\subsection{The Yukawa sector}
The Yukawa interactions for Standard Model and exotic quarks are given by 
\begin{align}
	\mathcal{L}^{Y}_{q,\ triplet}=\ &- \overline{Q}_m\brak{Y^d_{m\a} \eta^*d_{\a R}+Y_{m\a}^u\chi^*u_{m\a}}
	-\overline{Q}_3( Y^d_{3\a}\chi d_{\a R}+Y^u_{3\a}\eta u_{m\a})+\nonumber\\
	&-\overline{Q}_{m}(Y^J_{mn}\chi J_{nR})-\overline{Q}_{3}Y^J_{3}\chi J_{3R} +\textnormal{h.c.}\ , 
\end{align}
where $y^i_{d}$, $y^i_u$ and $y^i_E$ are the Yukawa couplings for down-,
up-type and exotic quarks, respectively. The masses of the exotic quarks are related to the vev of the neutral component of $\rho=(0,0,v_\rho)$ through the invariants
\begin{eqnarray}
	Q_1\,  \rho^*  D_R^*, Q_1\,  \rho^*  S_R^*&\sim & (\textbf{3},\textbf{3},-1/3)\times   (\textbf{1},\bar{\textbf{3}},-1)\times   (\bar{\textbf{3}},\textbf{1},4/3), \nonumber \\
	Q_3\,  \rho  T_R^* &\sim& (\textbf{3},\bar{\textbf{3}},2/3)\times   (\textbf{1},{\textbf{3}},1)\times   (\bar{\textbf{3}},\textbf{1},-5/3), 
\end{eqnarray}
responsible of the breaking $SU(3)_C\times   SU(3)_L\times   U(1)_X \to SU(3)_C\times   SU(2)_L\times   U(1)_Y$. It is clear that, since
$v_\rho\gg v_{\eta,\chi}$, the masses of the exotic quarks are
${\cal O}(\rm{TeV})$ whenever the relation $Y^J\sim \mathbb{1}$ is satisfied. 
In the lepton Yukawa sector, the interactions are more intricate due to the need to introduce an additional interaction with the scalar sextet to reproduce the physical lepton masses.
A typical Dirac mass term for leptons in the Standard Model is associated with the operator $\bar{l}_L H e_R$, where $l_L = (\nu_{eL}, e_L)$ represents the $SU(2)_L$ doublet. This has the representation content $(\bar{\textbf{2}}, 1/2) \times   (\textbf{2}, 1/2) \times   (\textbf{1}, -1)$ for $l_L$, $H$, and $e_R$, respectively, under the $SU(2)_L \times  U(1)_Y$ symmetry.
In the 331 model, however, both left- and right-handed lepton
components, e.g. $e_L$ and $e_R$, belong to the same multiplet. Therefore, to construct a $SO(1,3) \times  SU(3)_L$ singlet, two leptons from the same representation must be combined. This can be achieved, at least partially, through the operator
\begin{eqnarray}
	\mathcal{L}_{l,\, triplet}^{Yuk}&=& G^\eta_{a b}( l^i_{a \alpha}\epsilon^{\alpha \beta} l^j_{b \beta})\eta^{* k}\epsilon^{i j k} + \rm{h. c.}\nonumber\\
	&=& G^\eta_{a b}\, l^i_{a}\cdot l^j_{b}\,\eta^{* k}\epsilon^{i j k} + \rm{h. c.} \ , 
\end{eqnarray}
where the indices $a$ and $b$ run over the three generations of flavour, $\alpha$ and $\beta$ are Weyl indices, contracted to generate  an $SO(1,3)$ invariant ($l^i_{a}\cdot l^j_{b}\equiv l^i_{a \alpha}\epsilon^{\alpha \beta} l^j_{b \beta}$) from two Weyl fermions, and $i,j,k=1,2, 3$ are $SU(3)_L$ indices. 
The use of 
$\eta$ as a Higgs field is mandatory, since the components of the multiplet $l^j$ are $U(1)_X$ singlets. 
The representation content of the operator $l^i_a l^j_b$, according to $SU(3)_L$, is given by $\textbf{3}\times  \textbf{3}= \textbf{6} + \bar{\textbf{3}}$, with the 
$\bar{\textbf{3}}$ extracted by an anti-symmetrization over $i$ and $j$ via $\epsilon^{i j k}$. This allows to identify 
$l^i_a l^j_b \eta^{*k}\epsilon^{i j k}$ as an $SU(3)_L$ singlet. Considering that the two leptons are anticommuting Weyl spinors, and that the $\epsilon^{\alpha\beta}$ (Lorentz) and $\epsilon^{i j k}$ ($SU(3)_L$) contractions introduce two sign flips under the $a\leftrightarrow b$ exchange, the combination 
\begin{equation}
	M_{a b}=( l^i_{a}\cdot l^j_{b })\eta^{* k}\epsilon^{i j k} 
\end{equation}
is therefore antisymmetric under the exchange of the two flavours, implying
that even $G_{a b}$ has to be antisymmetric.  However, an antisymmetric $G^\eta_{a b}$ Yukawa matrix is not sufficient to provide mass to all the leptons. 
In fact, the diagonalization of $G^\eta$ by means of a unitary matrix 
$U$, namely $G^\eta=U \Lambda U^\dagger$, with $G^\eta$ antisymmetric in
flavour space, implies that its three eigenvalues are given by
$\Lambda=(0,\lambda_{22}, \lambda_{33})$, with $\lambda_{22}=-\lambda_{33}$,
i.e., one eigenvalue is null and the other two are equal in magnitude.
We shall solve this problem by introducing a second 
invariant operator, with the inclusion of a sextet $\sigma$,
\begin{equation}
	\sigma=\left(
	\renewcommand*{\arraystretch}{1.5}
	\begin{array}{ccc}
		\sigma_1^{++}&\sigma_1^+/\sqrt2&\sigma^0_1/\sqrt2\\
		\sigma_1^+/\sqrt2&\sigma_2^0&\sigma_2^-/\sqrt2\\
		\sigma^0_1/\sqrt2&\sigma_2^-/\sqrt2&\sigma_2^{--}
	\end{array}
	\right)\in(\textbf{1},\textbf{6},0),
\end{equation}
leading to the Yukawa term
\begin{equation}\label{lag}
	\mathcal{L}_{l, sextet}^{{Yuk.}}= G^\sigma_{a b} l^i_a\cdot l^j_b \sigma^*_{i,j},
\end{equation}
which allows to build a singlet out of the representation
$6$ of $SU(3)_L$, contained in $l^i_a\cdot l^j_b$, by combining it with
the flavour-symmetric
$\sigma^*$, i.e. $\bar{6}$.
Notice that $G^\sigma_{a b}$ is symmetric in flavour space. The
Yukawa Lagrangian in the lepton sector
can then be written as a combination of of triplet and sextet contributions:
\begin{equation}
	\mathcal{L}_{l}^{{Yuk.}}=\mathcal{L}_{l, sextet}^{{Yuk.}} + \mathcal{L}_{l, triplet}^{{Yuk.}} + \rm{h. c.}\ .
\end{equation}
It is interesting to note that, without considering the sextet, a doubly-charged scalar would not be able to decay into same-sign leptons. This is because,
if one had no sextet, the interaction responsible for the leptons
would only involve the scalar triplet $\eta$, which does not contain any
doubly-charged state. 

\section{Benchmark points}

In order to identify some viable benchmark points of the 331 model with $\beta = \sqrt{3}$, we follow a systematic procedure involving model implementation and phenomenological constraints. First, we construct the Lagrangian of the model using the \texttt{SARAH} package \cite{staubAutomaticCalculationSupersymmetric2011,staubSARAH4Tool2014}, which enables the automatic generation of the \texttt{SPheno} code \cite{porodSPhenoProgramCalculating2003,porodSPheno31Extensions2012}. This code is subsequently used to compute mass spectrum and relevant observables of the model.
To ensure consistency with current experimental data, we employ \texttt{HiggsBounds} \cite{bechtleHiggsBoundsConfrontingArbitrary2010,bechtleHiggsBounds200Confronting2011} and \texttt{HiggsSignals} \cite{bechtleHiggsSignalsConfrontingArbitrary2014} to test a possible benchmark point against the full set of Higgs sector constraints. These include exclusion limits from direct searches as well as precision measurements of the Higgs boson properties. Only the points that pass
all these constraints are considered as viable benchmark candidates. 

To streamline and automate the entire workflow, we employ the \texttt{SSP}
(\texttt{SARAH} Scan and Plot) package \cite{staubToolBoxImplementing2012}, which facilitates parameter space exploration in BSM scenarios. \texttt{SSP} allows for the automatic generation of SLHA input files for \texttt{SPheno}, execution of the spectrum calculation, and direct interfacing with \texttt{HiggsBounds} and \texttt{HiggsSignals} using the \texttt{SPheno} output. 
This automated pipeline enables a systematic and efficient scan over a predefined set of model parameters, verifying at each step whether the generated points are consistent with current Higgs data. Through this setup, one can perform comprehensive phenomenological analyses and identify viable regions of parameter space that satisfy both theoretical and experimental constraints. 

To ensure the consistency of the 331 model with the Standard Model (SM), the three vacuum expectation values (VEVs) must satisfy the hierarchical condition
	\begin{equation}
		v_\rho \gg v_\eta\,, \quad v_\rho \gg v_\chi \quad \text{and}\quad v_\rho \gg v_\sigma\ .
	\end{equation}
	These relations guarantee that the additional heavy particles predicted by the 331 model do not lie near the electroweak scale. Otherwise, such states would have already manifested in collider searches. 
	
	In order to reproduce the correct mass for the $W$ boson, the VEVs must also satisfy the electroweak constraint
	\begin{equation}
		v_\eta^2 + v_\chi^2 + v_\sigma^2 = v_W^2 = (246~\text{GeV})^2\ .
	\end{equation}
	This condition effectively reduces by one the number of independent parameters in the scalar potential. It is convenient to parametrize the three VEVs in terms of the Weinberg scale through two mixing angles $\beta_1$ and $\beta_2$ as follows:
	\begin{align}
		v_\eta &= v_W \sin\beta_2 \cos\beta_1\ , \\
		v_\chi &= v_W \sin\beta_2 \sin\beta_1\ , \\
		v_\sigma &= v_W \cos\beta_2\ ,
	\end{align}
	where we have imposed $\tan\beta_1 = 0.15$ and $\tan\beta_2 = 0.5$ for the scan of the parameter space.  
	
	The stability of the scalar potential restricts the possible values of the couplings. In particular, the self-couplings $\lambda_{\rho,\eta,\chi,\sigma}$ must be positive to guarantee the boundedness of the potential along the $\rho$, $\eta$, $\chi$, and $\sigma$ directions. Conversely, the mixed couplings $\lambda_{ij}$, $\zeta_{ij}$, $\xi_{ij}$, as well as the trilinear terms $f_{\rho\eta\chi}$ and $f_{\rho\sigma\chi}$, can in principle take negative values without violating stability, provided that the overall potential remains bounded from below.
	
	For the numerical scan, we fixed the following parameters:
	\begin{align}
		v_\rho &= 4~\text{TeV}\ , \\
		f_{\rho\eta\chi} &= f_{\rho\sigma\chi} = -4~\text{TeV}\ .
	\end{align}
	
	Given that the scalar sector of the 331 model contains a large number of free parameters (26 in total), we have adopted a stochastic exploration strategy. The allowed parameter space consistent with vacuum stability and with at least one scalar state identified as the observed Higgs boson is found to be extremely sparse due to the high dimensionality of the scalar potential and the interplay among couplings. Therefore the strategy that we adopted was to generate random points using the \texttt{SSP} framework within the following ranges:
	\begin{equation}
		0 < \lambda_{\rho,\eta,\chi,\sigma} < 3\ , \qquad \lambda_{ij},\, \zeta_{ij},\, \xi_{ij} \in [-1,\,1]\ ,
	\end{equation}
	which we have found to be the region with the major density of allowed points.
	We used \texttt{HiggsBounds} to test the compatibility of the selected benchmark points with the current experimental exclusion limits from Higgs searches. The \texttt{HiggsBounds} framework allows one to confront the predictions of extended Higgs sectors with a comprehensive collection of model-independent bounds derived from collider data.
	
	For each benchmark point, we provided as input masses, total widths and branching ratios of all neutral and charged Higgs bosons predicted by the 331 model, together with the relevant Higgs production cross sections normalized to their Standard Model (SM) equivalents. This allows \texttt{HiggsBounds} to internally compute the hadronic and partonic cross section ratios required for the comparison with the experimental analyses. The observables tested include both production and decay channels for neutral and charged Higgs bosons. Representative examples of these observables are \cite{bechtleHiggsBounds5TestingHiggs2020}:
	\begin{itemize}
		\item Neutral Higgs production via gluon fusion ($gg \to h_j$), bottom-quark associated production ($pp \to b\bar{b}h_j$), vector boson fusion ($pp \to qqh_j$), associated production with vector bosons ($pp \to Vh_j$, with $V=W,Z$), and top-associated production ($pp \to t\bar{t}h_j$);
		\item Higgs boson decays into Standard Model final states such as $h_j \to b\bar{b}$, $\tau^+\tau^-$, $\gamma\gamma$, $Z\gamma$, $WW$, and $ZZ$, as well as flavor-violating decays like $h_j \to \mu^\pm \tau^\mp$ or invisible final states;
		\item Non-standard decay topologies, including $h_k \to h_i h_j$, $h_j \to h_i Z$, and $h_j \to H_i^\pm W^\mp$, relevant for multi-Higgs models;
		\item Charged Higgs production processes such as $pp \to tbH^\pm$ and $pp \to H^\pm W^\mp$, with subsequent decays $H^\pm \to \tau^\pm \nu_\tau$, $H^\pm \to tb$, or $H^\pm \to W^\pm Z$;
		\item LHC searches in the $\tau^+\tau^-$, $b\bar{b}$, $WW$, $ZZ$, and $t\bar{t}$ final states, as well as resonant di-Higgs production ($pp \to \phi \to hh$) and cascade decays like $pp \to \phi_2 \to Z\phi_1$;
	\end{itemize}
        Only those benchmark points that passed all \texttt{HiggsBounds} exclusion tests were retained for further phenomenological investigation of exotic decay modes within the 331 mdel.
        In particular, we have selected the reference point in Table \ref{table:mass}. 
\begin{table}[h]
	\begin{center}
		\renewcommand{\arraystretch}{1.4}
		\begin{tabular}{|c|c|c|}
			\hline\hline
			\multicolumn{3}{|c|}{Benchmark Point}\\
			\hline
			\hline
			$m_{h_1}=125.3$ GeV & $m_{h_2}=6799.5$ GeV& $m_{h_3}=10489.8$ GeV\\
			\hline
			$m_{h_4}=17673.2$ GeV & $m_{h_5}=17737.2$ GeV& \\
			\hline
			$m_{Ah_1}=10489.5$ GeV& $m_{Ah_2}=17734.9$ GeV& $m_{H^\pm_1}=2566.4$ GeV\\
			\hline
			$m_{H^\pm_2}=10490.1$ GeV & $m_{H^\pm_3}=17673.5$ GeV &$m_{H^\pm_4}=17735.7$ GeV\\
			\hline
			$m_{H^{\pm\pm}_1}=10572.0$ GeV& $m_{H^{\pm\pm}_2}=17673.9$ GeV & $m_{H^{\pm\pm}_3}=17796.0$ GeV \\
			\hline
			$m_{Y^{\pm\pm}}=1325.7$ GeV& $m_{X^\pm}=1328.1$ GeV & $m_{Z'}=4150.4$ GeV\\
			\hline
			$m_D=1500.0$ GeV & $m_S=2000.0$ GeV& $m_T=3000.0$ GeV\\
			\hline
			\hline
		\end{tabular}
		\caption{Benchmark point for our collider study, consistent with the Standard Model Higgs mass and present exclusion limits on BSM physics.}\label{table:mass}
	\end{center}
\end{table}
For the purpose
of our investigation, a few comments on the
benchmark point in Table~\ref{table:mass} are in order.
The lightest neutral scalar $h_1$ has mass consistent with a SM Higgs boson,
while the mass of the lightest charged $H_1^\pm$ is about 2.5 TeV,
hence potentially in the reach of LHC. The other Higgs bosons are instead
too heavy to be searched for at LHC.
Consistently with the findings of \cite{Coriano:2020iiz},
doubly- ($Y^{\pm\pm}$) and singly-charged ($X^{\pm}$) vector bileptons have mass
about 1.3 TeV. Furthermore, the $Z'$ boson is slightly above 4 TeV and
the new heavy quarks feature masses $m_T\simeq 3$~TeV, $m_S\simeq 2$~TeV and
$m_D\simeq 1.5$~TeV, hence, at least from the kinematic viewpoint,
they are within the reach of LHC.

Additionally, within this parameter point of the scalar sector, we identify two benchmark points for the Yukawa couplings that are phenomenologically interesting and compatible with current collider bounds. These benchmarks can serve as reference points for further studies, including collider simulations and
investigation of flavour observables.
At this point, it is important to observe the the 331 model is sensible to the Yukawa couplings, since different choices can lead to
to the same CKM matrix, but with different phenomenology. This happens because in the Lagrangian of the 331 model the CKM matrix is not the only combination of quark rotation matrices, but they appear also singularly, as in the following interaction term:
\begin{equation}
	\mathcal{L}_{dTY}=\bar{T}_L\g^\m(U_L)_{j3}\textbf{d}_{j}Y_\m^{++}+\text{h.c.}
\end{equation}
where 
\begin{equation}
	\textbf{d}=\begin{pmatrix}
		d\\s\\b
	\end{pmatrix}
\end{equation}
The first benchmark that we have found is
\begin{equation}
	Y_d=\begin{pmatrix}
		2.03\times 10^{-4}	&	2.14\times 10^{-3}	&	2.26\times 10^{-2}	\\
		2.26\times 10^{-3}	&	1.77\times 10^{-2}	&	3.63\times 10^{-1}	\\
		2.26\times 10^{-1}	&	3.06	&	80.1
	\end{pmatrix},
\end{equation} 
which leads to the rotation matrix:
\begin{equation}
	U_{L}^d=\begin{pmatrix}
		-9.74\times 10^{-1}	&	2.27\times 10^{-1}	&	-4.21\times 10^{-3}	\\
		2.27\times 10^{-1}	&	9.74\times 10^{-1}	&	-3.01\times 10^{-2}	\\
		2.72\times 10^{-3}	&	3.03\times 10^{-2}	&	9.99\times 10^{-1}
	\end{pmatrix}.
\end{equation}
Likewise, the second one  
\begin{equation}
	Y_d=\begin{pmatrix}
		1.66\times 10^{-2}	&	8.37\times 10^{-5}	&	3.21\times 10^{-5}	\\
		4.27\times 10^{-4}	&	2.97\times 10^{-4}	&	9.76\times 10^{-5}	\\
		4.60\times 10^{-4}	&	4.08\times 10^{-4}	&	2.74\times 10^{-4}
	\end{pmatrix}
\end{equation}
leads to
\begin{equation}
	U_{L}^d=\begin{pmatrix}
		-1.68\times 10^{-3}	&	2.23\times 10^{-1}	&	-9.75\times 10^{-1}	\\
		-2.61\times 10^{-2}	&	9.75\times 10^{-1}	&	-2.23\times 10^{-1}	\\
		9.99\times 10^{-1}	&	2.59\times 10^{-2}	&	4.18\times 10^{-3}
	\end{pmatrix}
\end{equation}
In fact, these two benchmarks predict two different decay branching ratios for the exotic $T$ quark: the first one favours the decay into bottom quarks,
while in the second one the dominant mode is the $d$-quark channel.
\begin{figure}
	\centering
	\begin{tikzpicture}
		\begin{feynman}
			
			\vertex (dbar) at (0,1.5) {\(\bar{d}\)};
			\vertex (d) at (0,-1.5) {\(d\)};
			\vertex (v1) at (1.5,0);
			\vertex (v2) at (4,0);
			\vertex (T) at (5.8,1);
			\vertex (Tbar) at (5.8,-1);
			
			\vertex (Ypp) at (7.2,1.6);
			\vertex (bbar) at (7.5,0.7) {\(b\)};
			\vertex (lp1) at (9.2,2.2) {\(\ell^+\)};
			\vertex (lp2) at (9.2,1.0) {\(\ell^+\)};
			
			\vertex (Ymm) at (7.2,-1.6);
			\vertex (b) at (7.5,-0.7) {\(\bar{b}\)};
			\vertex (lm1) at (9.2,-1.0) {\(\ell^-\)};
			\vertex (lm2) at (9.2,-2.2) {\(\ell^-\)};
			
			\diagram* {
				(v1) -- [fermion] (dbar) ,
				(d)-- [fermion] (v1),
				(v1) -- [gluon, edge label=\(g\)] (v2),
				(v2) -- [fermion,edge label = \(T\)] (T),
				(v2) -- [anti fermion, edge label' = \(\bar{T}\)] (Tbar),
				
				(T) -- [fermion] (bbar),
				(T) -- [boson, edge label=\(Y^{++}\)] (Ypp),
				(Ypp) -- [anti fermion] (lp1),
				(Ypp) -- [anti fermion] (lp2),
				
				(Tbar) -- [anti fermion] (b),
				(Tbar) -- [boson, edge label'=\(Y^{--}\)] (Ymm),
				(Ymm) -- [fermion] (lm1),
				(Ymm) -- [fermion] (lm2),
			};
			
		\end{feynman}
	\end{tikzpicture}
	\caption{Example of $T\bar{T}$ pair production at LHC or FCC-$hh$,
          with $T$ decaying into a bilepton and a $b$ quark}
\end{figure}
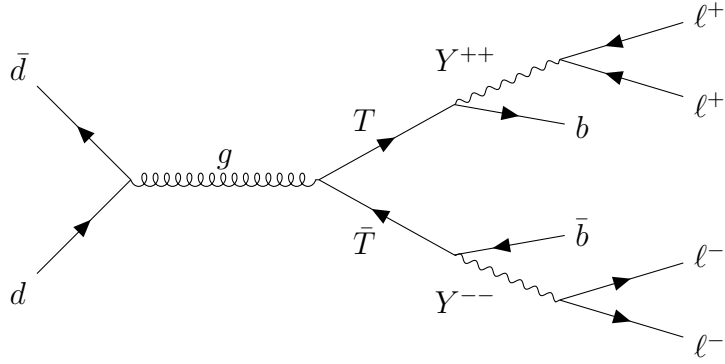\par
Before investigating the phenomenology of the bilepton model in the
representative point in Table~\ref{table:mass}, we wish to comment on the issue of
the perturbativity of the model. As discussed in \cite{diasConcerningLandauPole2005,martinezLandauPoleDecays2007} and pointed out in the
phenomenological analysis of \cite{corcellaExploringScalarVector2018},
the bilepton model possesses a 
a singular behaviour associated
with the theoretical embedding of $SU(2)$ in $SU(3)$ and matching conditions
between the couplings. In particular, the theory may lose its
perturbativity whenever the $U(1)$ coupling constant $g_X(\mu)$ is evaluated
at a scale such that $\sin^2\theta_W(\mu)\simeq 1/4$, which corresponds
to $\mu\simeq 4$~TeV. This singularity has often been labelled as
Landau pole, although, strictly speaking,
the Landau pole has a different meaning and
refers to the divergence in the QED coupling
at very high energy due to a positive renormalization-group
$\beta$-function \cite{Landau:1954nau}. More recently,
Ref.~\cite{Morisi:2025svd} found out that it is possible to further shift
the singularity of the bilepton model to higher energies
by adding a fourth family of both quarks and leptons.
As a matter of fact, the
present paper will deal with energy scales for bilepton production
well below 4 TeV, therefore we shall assume that we are still within
the perturbative regime and defer to future work any possible discussion
and comparison with \cite{Morisi:2025svd} 
on how to deal with such a singularity whenever the energy should approach
$\mu\simeq 4$~TeV.

\section{Phenomenology}
In this section, we investigate the phenomenology of heavy-quark production and their
decays into vector bileptons at LHC (14 TeV) and FCC-$hh$ (100 TeV).
In our benchmark point, heavy quarks like $T$ have mass 3 TeV: therefore, one can already
envisage that, as found in \cite{corcellaNonleptonicDecaysBileptons2022} for the purpose of
non-leptonic bilepton decays, the production cross section at LHC would be rather low, since
one may need at least a partonic centre-of-mass energy of $2m_T\simeq 6$~TeV.
Therefore, 
the expected number of events is quite suppressed even in the high-luminosity
phase. On the other hand, one may foresee a substantial number
of events at 100 TeV and integrated luminosity ${\cal L} = 3000$~fb$^{-1}$, as expected at
FCC-$hh$.

We focus on the primary production of $T\bar{T}$ pairs, followed by the decays $T \to Y^{++} b$ and $\bar{T} \to Y^{--} \bar{b}$, where the bileptons subsequently decay into same-sign muon pairs:
\begin{equation}
\label{ttbar}
pp \to T\bar{T} \to (Y^{++}b)(Y^{--}\bar{b}) \to (\mu^+ \mu^+ b)(\mu^- \mu^- \bar{b}).
\end{equation}
The final state at parton level thus consists of two same-sign muon pairs and two $b$ quarks, which hadronize through $b$-flavoured jets. 
For the chosen benchmark point (see Table~\ref{table:mass}), the heavy quark $T$ has a width of $\Gamma(T) \simeq 32.1$~GeV and the following branching ratios:
\begin{equation}
\text{BR}(T \to Y^{++} b) \simeq 47.1\%,\quad \text{BR}(T \to X^+ t) \simeq 46.3\%,\quad \text{BR}(T \to H_1^+ t) \simeq 4.6\%.
\end{equation}
Bileptons exhibit nearly democratic decay rates into all three lepton families:
\begin{equation}
\text{BR}(Y^{++} \to e^+ e^+) \simeq \text{BR}(Y^{++} \to \mu^+ \mu^+) \simeq \text{BR}(Y^{++} \to \tau^+ \tau^+) \simeq \frac{1}{3}.
\end{equation}
The hard scattering process is simulated at leading order (LO) using \texttt{MadGraph}~\cite{Alwall:2011uj}, matched to \texttt{HERWIG 6}~\cite{Corcella:2000bw} for parton showering, hadronization and underlying event modelling. The possible inclusion of higher-order corrections to the
hard scattering, e.g. NLO,
as well as detector effects, is deferred to future work. 

The LO cross section for the signal process~\eqref{ttbar} at $\sqrt{s} = 14$~TeV is approximately $\sigma({\rm LHC}) \simeq 6 \times 10^{-7}$~pb, rendering observation unfeasible even with the full HL-LHC dataset. At $\sqrt{s} = 100$~TeV, however, the cross section increases to $\sigma({\rm FCC}) \simeq 0.02$~fb, making the signal potentially observable at FCC-$hh$.

The dominant Standard Model backgrounds to the signal in Eq.~\eqref{ttbar} include:
\begin{itemize}
    \item $ZZ$ production in association with a $b\bar{b}$ pair, where both $Z$ bosons decay into $\mu^+\mu^-$ (background $b_1$), i.e.
    \begin{equation}
    \label{back1}
    pp \to ZZb\bar{b} \to (\mu^+\mu^-)(\mu^+\mu^-) b\bar{b}; 
    \end{equation}
    \item associated production of $t\bar{t}Z$, with the $Z$ decaying into $\mu^+\mu^-$, and both top quarks decaying leptonically (background $b_2$), namely
    \begin{equation}
    \label{back2}
    pp \to t\bar{t}Z \to (b\mu^+\nu_\mu)(\bar{b} \mu^- \bar{\nu}_\mu)(\mu^+\mu^-).
    \end{equation}
\end{itemize}
The $ZZb\bar b$ background yields the same final state as our signal, however,
the kinematics of the particles in the final state are expected to be
rather different.
In fact, $ZZb\bar b$ processes yield lepton pairs with different sign and invariant mass about
$m_Z\simeq 91.19$~GeV, while the signal produces
same-sign lepton pairs with mass around
$m_Y\simeq 1.3$~TeV.
As for the $t\bar t Z$ background, its final states are different from the signal because of the
presence of neutrinos, however, as already observed in \cite{corcellaBileptonSignaturesLHC2017,corcellaNonleptonicDecaysBileptons2022}, processes
with top quarks in the dilepton channel can mimic the signal whenever the
missing energy due to neutrinos is small.

In the following, we simulate all backgrounds by means of \texttt{MadGraph},
matched to HERWIG 6, as was done for the signal, and concentrate ourselves on
FCC-$hh$.
The following LO cross sections are obtained at $\sqrt{s} = 100$~TeV:
\begin{equation}
\sigma(ZZb\bar{b}) \simeq 4.7 \times 10^{-3}~{\rm pb}, \qquad \sigma(t\bar{t}Z) \simeq 1.7 \times 10^{-2}~{\rm pb}.
\end{equation}

In line with Ref.~\cite{corcellaNonleptonicDecaysBileptons2022}, we apply the following acceptance cuts to transverse momentum ($p_T$),
pseudorapidity ($\eta$) and invariant opening angle
$\Delta R=\sqrt{(\Delta\eta)^2+(\Delta\phi)^2}$, where $\Delta\eta$ 
and $\Delta\phi$ are azimuthal and pseudorapidity gaps of 
final-state muons and $b$-jets, as well as to the missing transvere energy
(MET)\footnote{As in Ref.~\cite{corcellaNonleptonicDecaysBileptons2022}
the missing transverse energy is defined as ${\rm MET}=\sqrt{\left(\sum_{i=\nu}p_{x,i}\right)^2+  \left(\sum_{i=\nu}p_{y,i}\right)^2}$.}: 
\begin{eqnarray}
\label{cuts}
&& p_{T,b} > 30~{\rm GeV},\quad p_{T,\mu} > 20~{\rm GeV},\quad |\eta_b| < 4.5,\quad |\eta_\mu| < 2.5, \nonumber \\
&& \Delta R_{bb} > 0.4,\quad \Delta R_{\mu\mu} > 0.1,\quad \Delta R_{b\mu} > 0.4,\quad {\rm MET} < 200~{\rm GeV}.
\end{eqnarray}
As in our previous work, these cuts reflect a conservative implementation of the lepton–jet overlap-removal algorithm used in LHC analyses~\cite{ATLAS:2018zzq}, and the MET requirement targets background~\eqref{back2}, which includes final-state neutrinos. However, the MET cut is applied universally to account for neutrinos from hadron decays in all processes.
To ensure realistic projections, we incorporate $b$-tagging efficiency. Although the efficiency depends on the $b$-jet kinematic properties, we adopt a flat approximation, as done in~\cite{corcellaNonleptonicDecaysBileptons2022}, and defer a detailed modeling to future work. Following Ref.~\cite{selvaggi}, we assume a tagging efficiency of $\epsilon_b = 0.82$\footnote{This efficiency applies to $b$-jets with pseudorapidity $|\eta_j| < 2.5$ and transverse momentum in the range $10~{\rm GeV} < p_{T,j} < 500~{\rm GeV}$, which covers the dominant phase space of the events considered.}.

After applying the cuts in Eq.~\eqref{cuts} and accounting for $b$-tagging, the expected number of signal events at FCC-$hh$ with ${\cal L} = 3000$~fb$^{-1}$ is approximately $N(2b4\mu)_s \simeq 32880$
\footnote{We also varied the $T$ mass between 2 and 5 TeV, and found an expected
number events at FCC-$hh$ between 9750 and 60450. The HL-LHC cross sections are instead
between $2\times 10^{-6}$~pb and $2\times 10^{-9}$~pb, hence too low to see any signal.}.
The corresponding background event yields are $N(2b4\mu)_{b_1} \simeq 1288$ for $ZZb\bar{b}$ and $N(2b4\mu)_{b_2} \simeq 8338$ for $t\bar{t}Z$.

Strictly speaking, one should also consider processes where $Z$ bosons are
accompanied
by light jets which are misidentified as $b$-jets, but one can anticipate that,
after the cuts in Eq.~(\ref{cuts}) and mistag efficiencies are
applied \footnote{In Ref.~\cite{corcellaNonleptonicDecaysBileptons2022},
an average mistag rate $\epsilon_j=0.15$ was used.},
only few dozens event survive, which
makes such a background negligible when searching for heavy quarks and
bileptons at both LHC and FCC-$hh$.
Furthermore, unlike Ref.~\cite{corcellaNonleptonicDecaysBileptons2022}, we
do not include the 4-top
Standard Model background, namely $pp\to (t\bar t)(t\bar t)$ followed by
$t\to bW^+$ and $W^+\to \mu^+\nu_\mu$. In fact, this process would yield a final state with 
4 $b$-jets, while our signal exhibits only  2 $b$-jets.
\footnote{In principle, one may even have 2 $b$-jets
starting from 4 $b$-quarks, e.g., whenever two pairs of
$B$-hadron are very close in the
phase space and get clustered in the same jet, but nevertheless,
as long as one uses a reasonable jet algorithm, such events are very
much suppressed.}.

\begin{figure}[b]
\centerline{\resizebox{0.49\textwidth}{!}
{\includegraphics{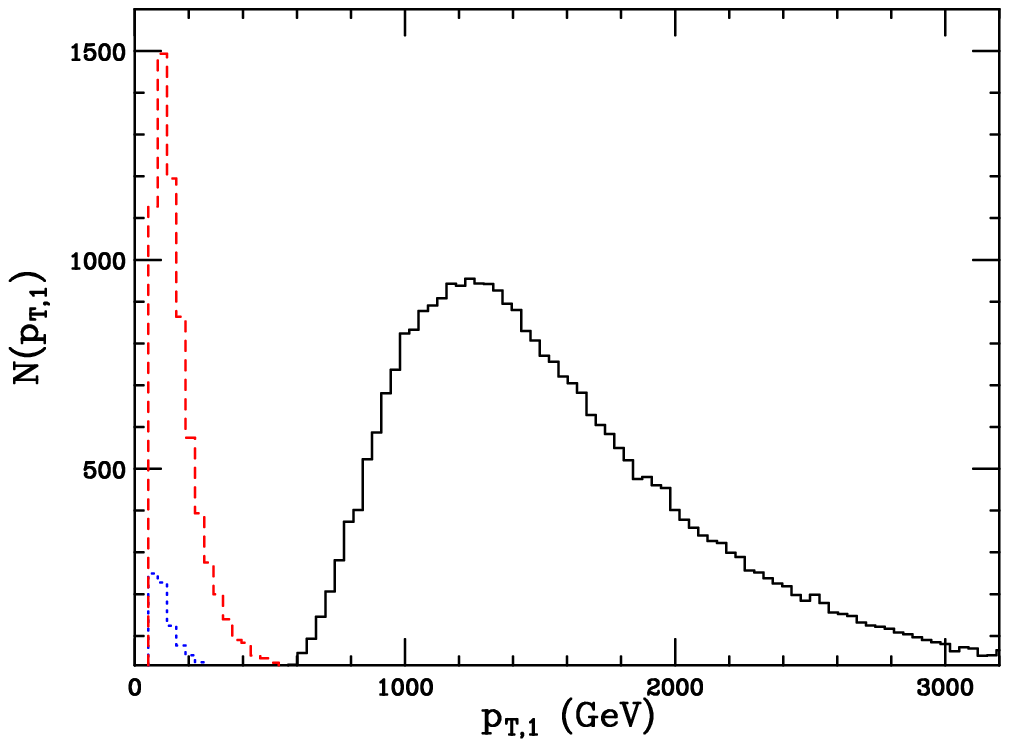}}%
\hfill%
\resizebox{0.49\textwidth}{!}{\includegraphics{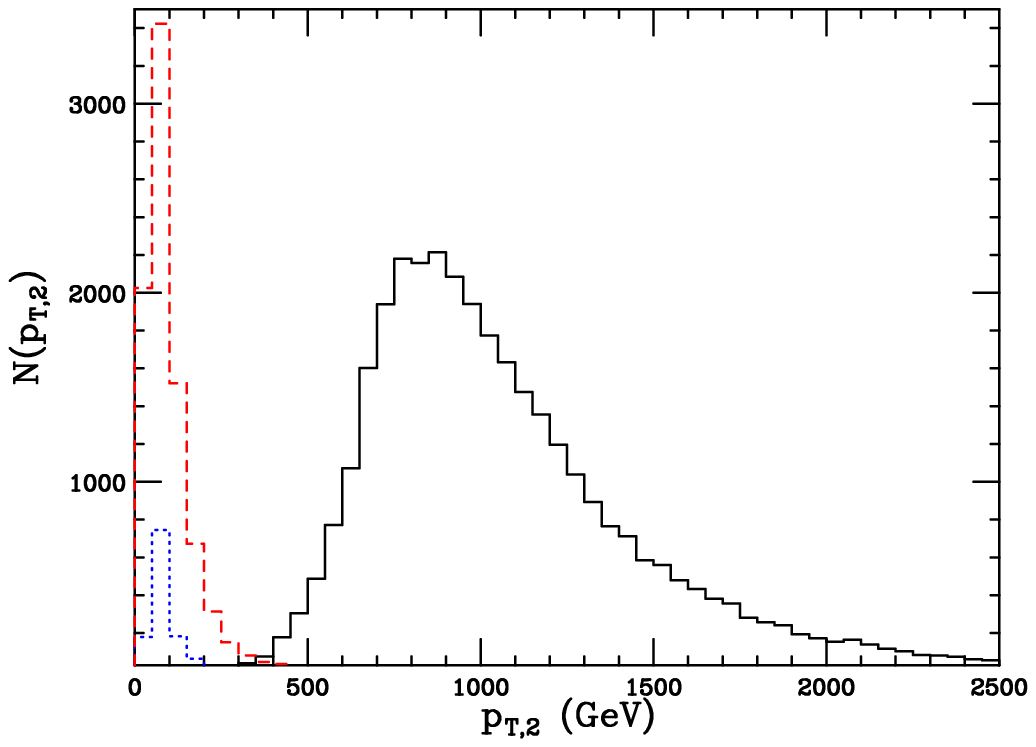}}}
\caption{Transverse momentum of the hardest (left) and next-to-hardest
  muon (right) for the signal (solid) and the backgrounds $b_1$ (dots) and
  $b_2$ (dashes).}
\label{ptl}
\end{figure}

As in previous work, we present some distributions according to
our bilepton signal in Eq.~(\ref{ttbar}) and backgrounds
Eqs.~(\ref{back1}) and (\ref{back2}).
All spectra are normalized in such a way that $N(x)$ yields the number
of expected events in the histogram bin around $x$ at FCC-$hh$. 
In Fig.~\ref{ptl}, we present the transverse momentum of the hardest
($p_{T,1}$, left) and next-to-hardest ($p_{T,2}$, right) muon
for the signal (solid) and backgrounds $b_1$ (dots) and $b_2$ (dashes):
overall, signal and background distributions look very different, which makes
their separation potentially straightforward.
The signal starts to be substantial at $p_{T,1}\simeq 600$~GeV and
$p_{T,2}\simeq 400$~GeV, peaks around 1.2 TeV ($p_{T,1}$) and
800 GeV ($p_{T,1}$) and vanishes for $p_{T,1}>3.5$~TeV and 
$p_{T,2}>2.5$~TeV. Background distributions are instead quite sharp:
$b_1$ is substantial only for $p_{T,1}< 200$~GeV and $p_{T,2}< 200$~GeV
and is maximum at low $p_{T,1}$ and $p_{T,2}$; $b_2$ yields a substantial number
of events for $p_{T,1}< 500$~GeV and $p_{T,2}< 500$~GeV
and is very sharply peaked
about 200 GeV ($p_{T,1}$) and 100 GeV ($p_{T,2}$).
Our signal features the production of heavy $T\bar T$ pairs and therefore the partonic centre-of mass energy is typically ${\cal O}(10~{\rm TeV})$
, which leads to bileptons and hence hardest leptons with transverse
momentum about
$p_T\sim {\cal O}(1~{\rm TeV})$. On the other hand, the $ZZb\bar b$ and
$t\bar tZ$ backgrounds are characterized by partonic centre-of-mass
energies which
are typically one order of magnitude lower and therefore the final-state
leptons are much softer than those originating from bileptons.

\begin{figure}
\centerline{\resizebox{0.49\textwidth}{!}
{\includegraphics{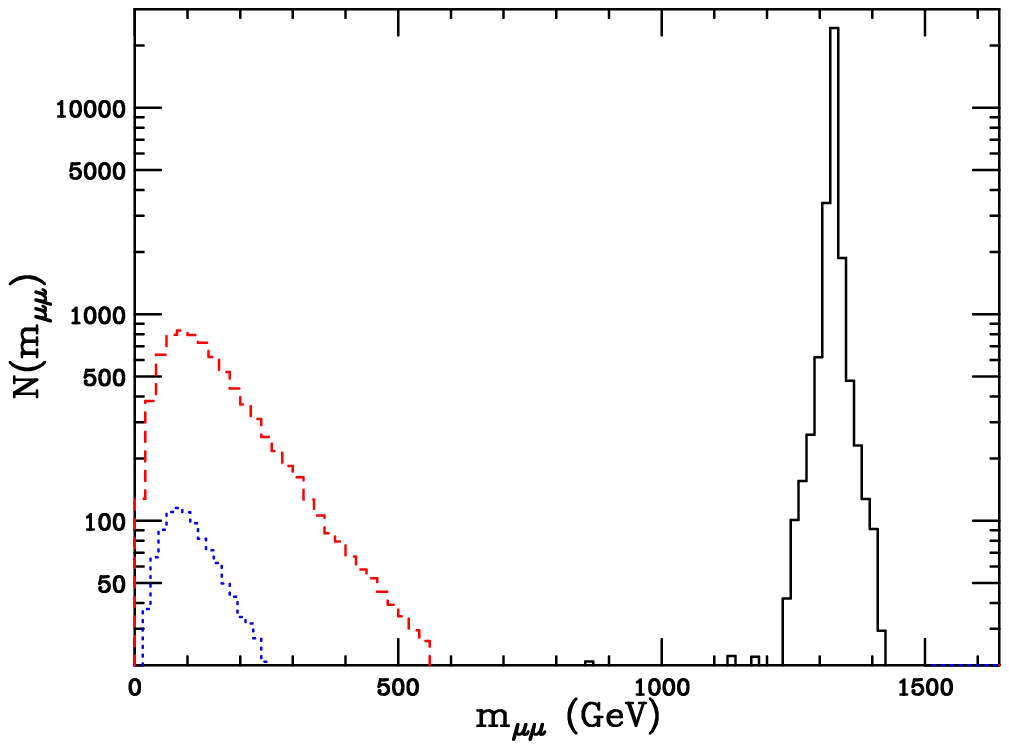}}%
\hfill%
\resizebox{0.49\textwidth}{!}{\includegraphics{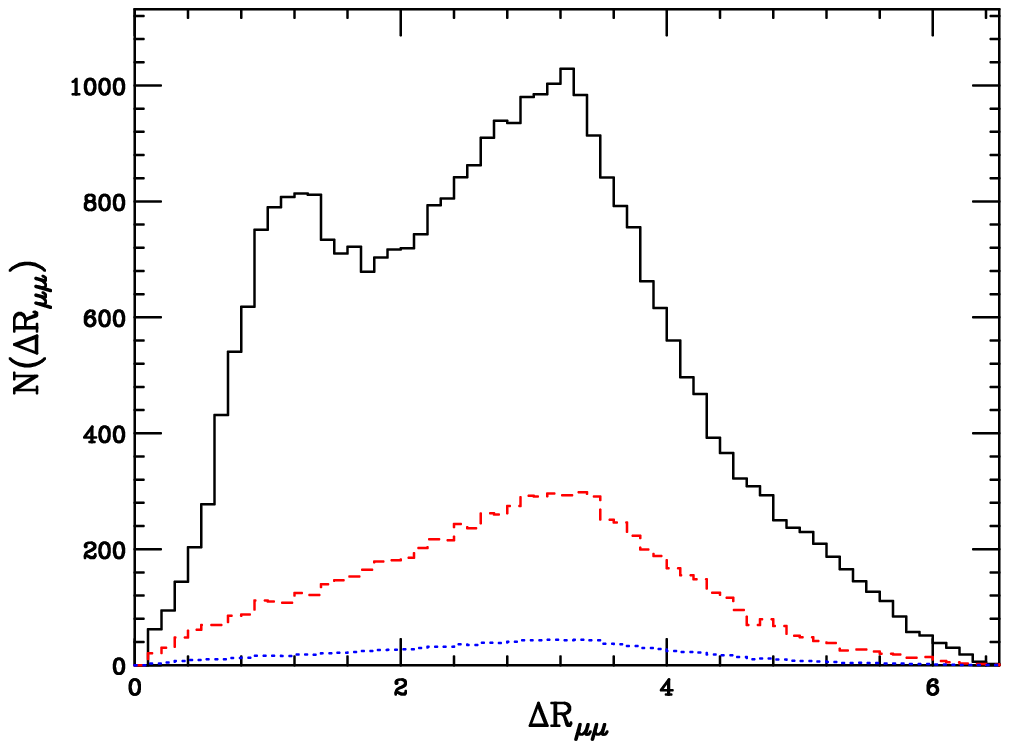}}}
\caption{Invariant-mass distribution (left) and invariant opening angle
  (right) between the two hardest
muons, according to the bilepton signal and the SM backgrounds . The histograms are labelled as in Fig.~\ref{ptl}.}
\label{mlldr}
\end{figure}

Fig.~\ref{mlldr} shows the invariant mass distributions $m_{\mu\mu}$ of same-sign muon pairs (left) and the invariant opening angle $\Delta R_{\mu\mu}$ 
  between the two hardest muons (right).
As observed in previous work as well, the invariant mass distribution yielded by the
bilepton model is strikingly different from the others, since it is sharply peaked
at the bilepton mass, i.e. $m_{\mu\mu}\simeq m_Y\simeq 1.3$~TeV; the background spectra 
are instead broader and peaked at a lower $m_{\mu\mu}\simeq$~80-90 GeV.
In fact, as discussed above, the backgrounds typically feature much softer leptons: furthermore, in order to get two leptons with the same sign, one
needs to pair the decay products of either two $Z$ bosons or
one $Z$ and one $W$ from top decays, which may
be quite distant in the phase space. Such effects are therefore
responsible of the
fact that the background $m_{\mu\mu}$  spectra are much softer and broader than
the signal.

As for $\Delta R_{\mu\mu}$, the bilepton signal is remarkably above the
background over the whole range and exhibits two maxima about
$\Delta R_{\mu\mu}\simeq 1$ and $\Delta R_{\mu\mu}\simeq 3$, with a dip in the
middle. Both backgrounds $b_1$ and $b_2$ increase until they reach a peak
value around $\Delta R_{\mu\mu}\simeq 3$ and rapidly vanish
for $\Delta R_{\mu\mu}>6$. All distributions are rather broad, 
which means that the two hardest leptons come either from the same
bilepton or vector boson (small $\Delta R$) or from two different ones (large $\Delta R$).

\begin{figure}
\centerline{\resizebox{0.49\textwidth}{!}
{\includegraphics{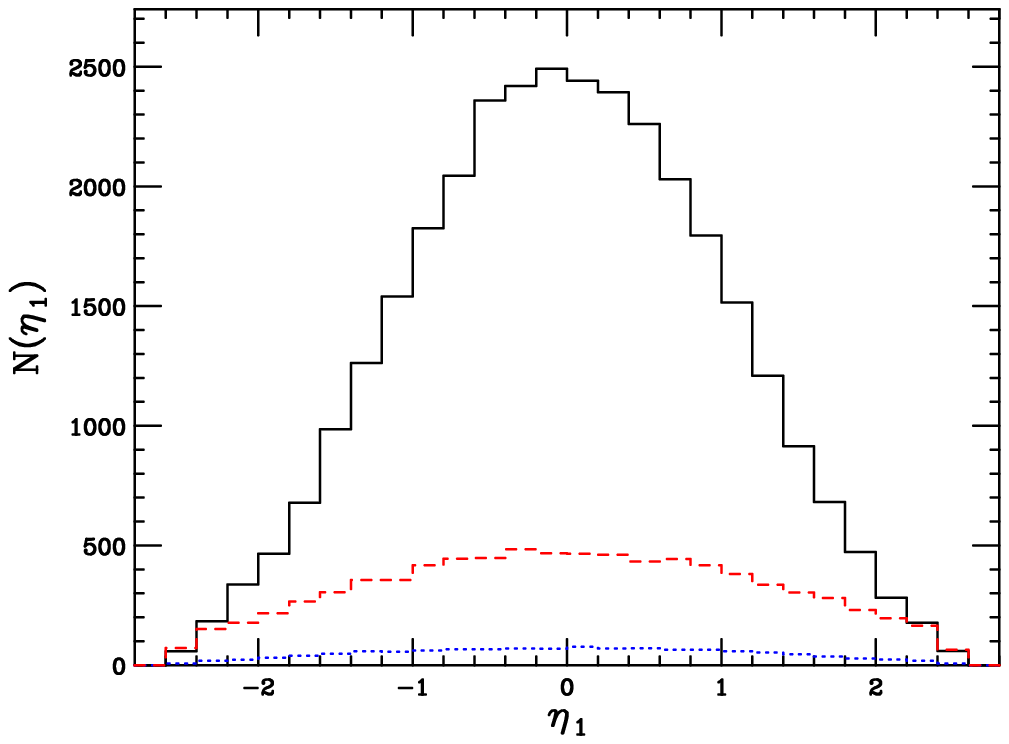}}%
\hfill%
\resizebox{0.49\textwidth}{!}{\includegraphics{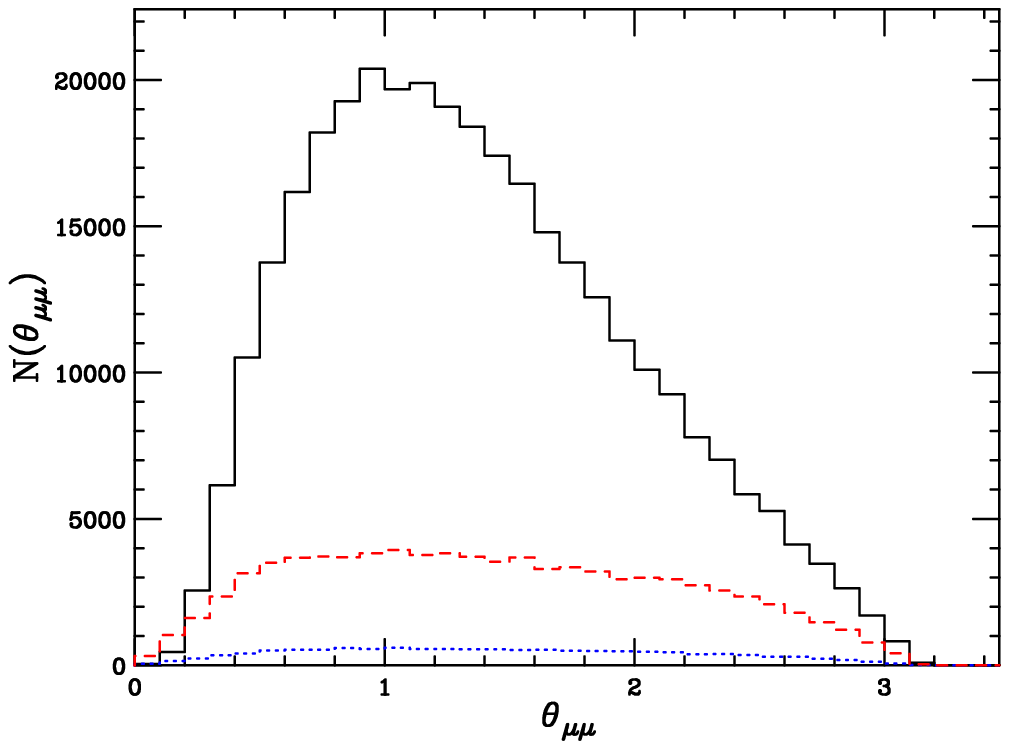}}}
\caption{As in Fig.~\ref{ptl}, but displaying the pseudorapidity of the
  hardest lepton (left) of the angle (right) between the two hardest ones.}
\label{etath}
\end{figure}

Fig.~\ref{etath} presents the pseudorapidity of the hardest muon $\eta_1$
and the angle $\theta_{\mu\mu}$ between the two hardest ones.
The pseudorapidity spectra exhibit a rise-and-fall behaviour, peaked at
$\eta_1\simeq 0$, which means that in both signals and backgrounds the
production of the hardest lepton in the central-rapidity region dominates.
The difference between the distributions is mostly due to normalization
and to the largest number of events yielded at FCC-$hh$ by the signal.
One can also notice that the $ZZb\bar b$ pseudorapidity is rather
flat over the full $-3<\eta_1<3$ range.

\begin{figure}
\centerline{\resizebox{0.49\textwidth}{!}
{\includegraphics{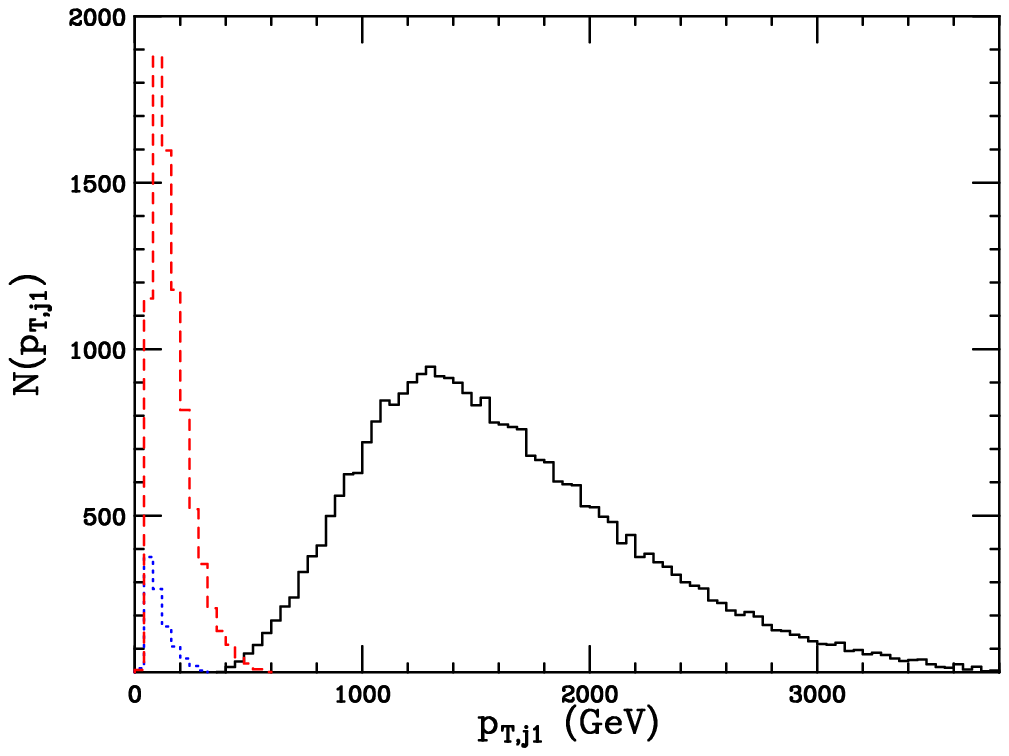}}%
\hfill%
\resizebox{0.49\textwidth}{!}{\includegraphics{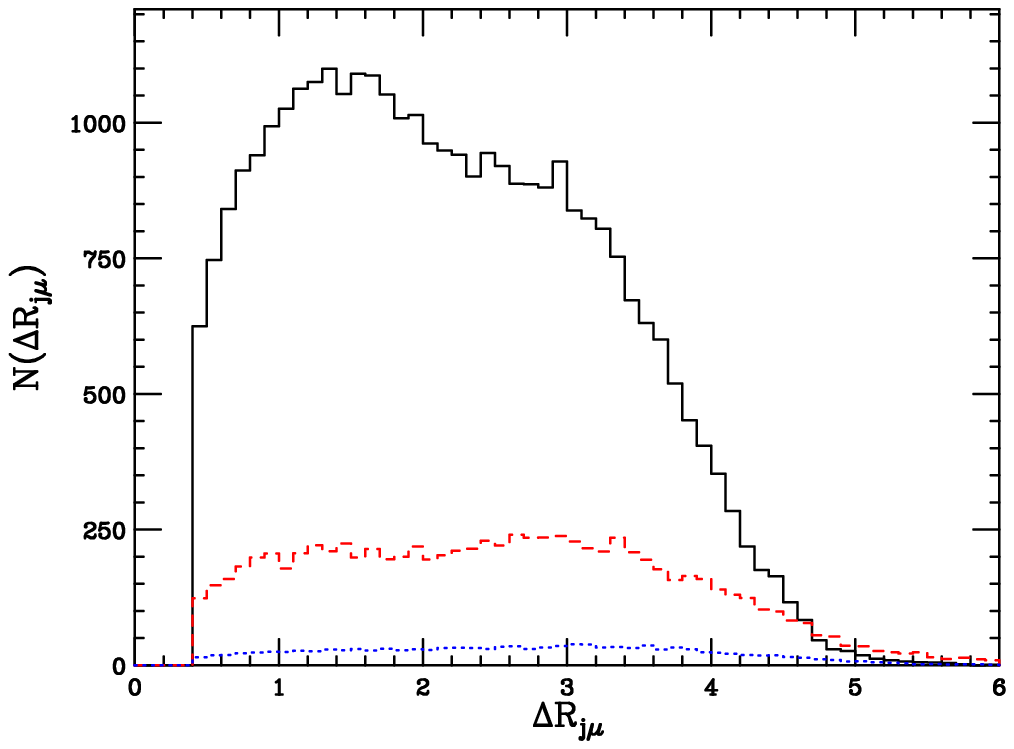}}}
\caption{Transverse momentum of the hardest jet (left) and invariant opening angle
  among hardest jet and hardest muon. Histograms are labelled like in the
previous figures.}
\label{ptjdr}
\end{figure}

Similar observations hold also for $\theta_{\mu\mu}$: the signal is pretty broad and
peaked at $\theta_{\mu\mu}\simeq 1$, in agreement with the finding of
a large number of events in the central pseudorapidity region, 
while the backgrounds are quite flat.
Finally, Fig.~\ref{ptjdr} shows distributions which involve the $b$-jets 
accompanying the four muons as well: the transverse momentum of the hardest jet $p_{T,j1}$
(left) and $\Delta R_{j\mu}$ between hardest jet and hardest muon
(right). 
The $p_{T,j1}$ spectra are similar to the leptonic ones in Fig.~\ref{ptl}, with the
signal predicting jets at large transverse momenta and background jets concentrated
in the low-$p_T$ range. In fact, $b$-quarks and $b$-jets produced in
$T\to Y^{++}b$, where the $T$ mass is about 3 TeV,
are typically much harder that those in $ZZb\bar b$ or top decays in $t\bar tZ$.
As for the $\Delta R_{j\mu}$ predictions, they mostly differ because of the
normalization: the signal spectrum is broad and peaked about
$\Delta R_{j\mu}\simeq 1.5$,
while the backgrounds are flatter, with $b_2$ yielding even more events
at large $\Delta R_{j\mu}>4.6$. As observed in the lepton case, the
broadness of such spectra are due to the fact that the hardest jet and the
hardest lepton can come from either the same (small $\Delta R$) or
different decay chains (large $\Delta R$).

Before concluding this section, as in
\cite{corcellaExploringScalarVector2018,corcellaNonleptonicDecaysBileptons2022}
we wish to calculate the significance $s$ to discriminate the bilepton signals $S$ through heavy quark
decays from the overall background $B$
\begin{equation}\label{signif}
s=\frac{S}{\sqrt{B+\sigma_B^2}}.
\end{equation}
In (\ref{signif}), $\sigma_B$ is the systematic error on $B$, which
we roughly estimate as $\sigma_B\simeq 0.1 B$.
We then obtain $s\simeq 34$, which underlines, once again, the sensitivity of
FCC to heavy quarks and bileptons.

\section{Conclusions}
As the LHC experiments have so far revealed no evidence of physics beyond the Standard Model, we revisited the 331 model, which predicts a variety of new particles, including bileptons, i.e. gauge bosons with charge and lepton number
$\pm 2$, heavy quarks with electric charges $5/3$ and $4/3$, a $Z'$ boson, and an extended Higgs sector. Our analysis focused on the production of heavy quarks $T$ with charge $5/3$, which decay into bileptons and bottom quarks:
the production of vector bileptons is in fact
a distinctive feature of the 331 model.
This process results in final states characterized by two $b$-jets and two same-sign lepton pairs, with the leptons arising from the bilepton decays.

In our investigation, we focused on heavy quarks $T$ with
mass around 3 TeV and chose a benchmark point consistent with current
exclusion limits and theoretical expectations for the bilepton mass, i.e. 
about 1.3~TeV, in agreement with the finding of \cite{Coriano:2020iiz}.
In the considered scenario, bileptons are assumed to
decay into same-sign muon pairs.
As for the backgrounds, we explored $ZZb\bar b$ and $t\bar tZ$ production,
with the $Z$ bosons decaying into muon pairs and dilepton top decays, i.e.
$t\to bW^+$ and $W^+\to \mu^+\nu_\mu$, with
small values of the missing energy due to neutrinos.

Under these assumptions, we investigated a number of observables related to
both leptons and jets in the final states, i.e. muon and $b$-jet
transverse momenta, pseudorapidities, production polar angles,
invariant opening angles $\Delta R$, as well as the invariant mass
of same-sign muons.
We generally found that the bilepton signal can be cleanly separated from the Standard Model backgrounds and should be detectable with very high significance at a future 100~TeV hadron collider, such as the FCC-$hh$.
In fact, the kinematic features of leptons and $b$-jets originating from decays
of a 3 TeV heavy quark and a bilepton with mass 1.3 TeV are pretty much
different from those in Standard Model vector boson and top decays.
Unlike FCC-$hh$, the expected cross sections and event yields at the LHC,
even with the projected high-luminosity upgrade, are too low to allow any meaningful detection and therefore our analysis was just concentrated on
FCC-$hh$.

In summary, we argue that the bilepton signal explored in this work — together with the scenarios proposed in our previous studies~\cite{corcellaExploringScalarVector2018,corcellaNonleptonicDecaysBileptons2022} — represents a compelling case for investigation at future collider experiments, which may shed light
on physics Beyond the Standard Model.


\bibliographystyle{jhep}
\bibliography{tdec}

\end{document}